\documentclass[aps,nofootinbib,showpacs,showkeys,preprint
] {revtex4}

\usepackage{epsf,epsfig,subfigure,axodraw,graphicx,amsmath,amssymb}
\usepackage{color}

\def\lsim{\lower.7ex\hbox{$\;\stackrel{\textstyle<}{\sim}\;$}}

\def\Qem{{$Q_{\rm em}$}}

 \def\Z{{\bf Z}}

\def\EE{E$_8\times$E$_8^\prime$}

 \def\Z{{\bf Z}}

\def\EE{E$_8\times$E$_8^\prime$}
\def\Eo{E$_8$}
\def\Eh{E$_8'$}

\def\fourb{\overline{\bf 4}}
\def\four{{\bf 4}}
\def\two{{\bf 2}}
\def\fiveb{\overline{\bf 5}}
\def\five{{\bf 5}}
\def\one{{\bf 1}}
\def\six{{\bf 6}}
\def\threeb{\overline{\bf 3}}
\def\three{{\bf 3}}

\begin{document}
\preprint{SNUTP 07-007}
\title{ Gauge mediated supersymmetry breaking without exotics
 in orbifold compactification}
\author{Jihn E.  Kim 
}
\address{Department of Physics and Astronomy and Center for Theoretical
 Physics, Seoul National University, Seoul 151-747, Korea }

\begin{abstract}
 {We suggest SU(5)$'$ in the hidden sector toward a possible
gauge mediated supersymmetry breaking scenario for removing the SUSY
flavor problem, with an example constructed in $\Z_{12-I}$ with
three families. The example we present has the Pati-Salam type
classification of particles in the observable sector and has no
exotics at low energy. We point out that  six or seven very light
pairs of ${\bf 5}'$ and $\overline{\bf 5}'$ out of ten vectorlike
$\five'$ and $\fiveb'$ pairs of SU(5)$'$ is achievable, leading to a
possibility of an unstable supersymmetry breaking vacuum. The
possibility of different compactification radii of three two tori
toward achieving the needed coupling strength is also suggested.
  }
\end{abstract}


 \maketitle

\section{Introduction}

The gauge mediated supersymmetry breaking (GMSB) has been proposed
toward removing the SUSY flavor problem \cite{DineNeslson}. However,
there has not appeared yet any satisfactory GMSB model from
superstring compactification, satisfying all phenomenological
constraints.

The GMSB relies on dynamical supersymmetry breaking \cite{Shirman}.
The well-known GMSB models are an SO(10)$'$ model with $\bf 16'$ or
$\bf 16'+10'$ \cite{SO10}, and an SU(5)$'$ model with $\bf
10'+\overline{5}'$ \cite{ADS}. If we consider a metastable vacuum
also, a SUSY QCD type is possible in SU(5)$'$ with six or seven
flavors, satisfying $N_c+1\le N_f<\frac32 N_c$ \cite{ISS}. Three
family standard models (SMs) with this kind of hidden sector are
rare. In this regard, we note that the flipped SU(5) model of Ref.
\cite{KimKyaegut} has one {\bf 16}$'$ and one {\bf 10}$'$ of
SO(10)$'$, which therefore can lead to a GMSB model. But as it
stands, the confining scale of SO(10)$'$ is near the GUT scale and
one has to break the group SO(10)$'$ by vacuum expectation values of
{\bf 10}$'$ and/or {\bf 16}$'$. Then, we do not obtain the spectrum
needed for a GMSB scenario and go back to the gaugino condensation
idea. If the hidden sector gauge group is smaller than SU(5)$'$,
then it is not known which representation necessarily leads to SUSY
breaking. The main problem in realizing a GMSB model is the
difficulty of obtaining the supersymmetry (SUSY) breaking confining
group with appropriate representations in the hidden sector while
obtaining a supersymmetric standard model (SSM) with at least three
families of the SM in the observable sector.

In this paper, we would like to address the GMSB in the orbifold
compactification of the \EE\ heterotic string with three families at
low energy.  A typical recent example for the GMSB is
 $$W= m\overline{Q}Q+\frac{\lambda}{M_{Pl}}Q\overline{Q}f
 \bar f+Mf\bar f$$
where $Q$ is a hidden sector quark and $f$ is a messenger. Before
Intriligator, Seiberg and Shih (ISS) \cite{ISS}, the GMSB problem
has been studied in string models \cite{GMSBst}. After \cite{ISS}
due to opening of new possibilities, the GMSB study has exploded
considerably and it is known that the above idea is easily
implementable in the ISS type models \cite{Kitano}. Here, we will
pay attention to the SUSY breaking sector, not discussing the
messenger sector explicitly. The messenger sector $\{f,\cdots\}$ can
be usually incorporated, using some recent ideas of \cite{Kitano},
since there appear many heavy charged particles at the GUT scale
from string compactifications. The three family condition works as a
strong constraint in the search of the hidden sector
representations.

In addition, the GUT scale problem that the GUT scale is somewhat
lower than the string scale is analyzed in connection with the GMSB.
Toward the GUT scale problem, we attempt to introduce {\it two
scales of compactification in the orbifold geometry}. In this setup,
we discuss physics related to the hidden sector, in particular the
hidden sector confining scale related to the GMSB. If the GMSB scale
is of order $ 10^{13}$ GeV, then the SUSY breaking contributions
from the gravity mediation and gauge mediation are of the same order
and the SUSY flavor problem remains unsolved. To solve the SUSY
flavor problem by the GMSB, we require two conditions: one is the
{\it relatively low hidden sector confining scale} ($<10^{12}$ GeV)
and the other is the {\it matter spectrum allowing SUSY breaking}.

Toward this kind of GMSB, at the GUT scale we naively expect a {\it
smaller coupling constant for} a relatively big {\it hidden sector
nonabelian gauge group} (such as SU(5)$'$ or SO(10)$'$) than the
coupling constant of the observable sector. But this may not be
needed always.

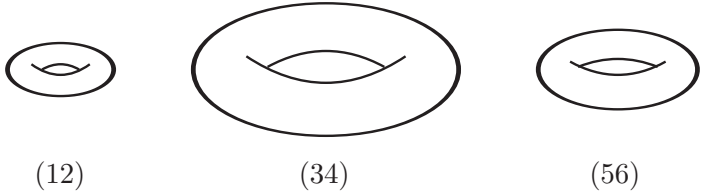
\begin{figure}[t]
\begin{center}
\begin{picture}(400,60)(0,0)
\SetWidth{0.9}
 \Oval(70,35)(10,20)(0) \Curve{(59,37)(70,33)(81,37)}
 \Curve{(63,35)(70,37)(77,35)}
 \Oval(280,35)(15,30)(0)\Curve{(262,38)(280,33)(298,38)}
   \Curve{(265,36)(280,39)(295,36)}
 \Oval(170,35)(25,50)(0) \Curve{(140,40)(170,30)(200,40)}
    \Curve{(148,36)(170,42)(192,36)}

\Text(70,-5)[c]{(12)}\Text(170,-5)[c]{(34)}\Text(280,-5)[c]{(56)}
\end{picture}
\caption{Radii of three tori can be different.}\label{fig:Thtorii}
\end{center}
\end{figure}

The radii of three two tori can be different in principle as
depicted in Fig. \ref{fig:Thtorii}.
For simplicity, we assume the same radius $r$ for (12)- and (56)-
tori. A much larger radius $R$ is assumed for the second (34)-torus.
For the scale much larger than $R$, we have a 4D theory. In this
case, we have four distance scales, $ R, r, \alpha'=M_s^{-2},\ {\rm
and}\ \kappa=M_P^{-1}, $ where $\alpha'$ is the string tension and
$M_P$ is the reduced Planck mass. The Planck mass is related to the
compactification scales by $M_P^2\propto M_s^8 r^4 R^2$. Assuming
that strings are placed in the compactified volume, we have a
hierarchy $\frac1R<\frac1r<M_s<M_P$. The customary definition of the
GUT scale, $M_{\rm GUT}$, is the unification scale of the QCD and
electroweak couplings.

For the 4D calculation of the unification of gauge couplings to make
sense, we assume that the GUT scale is below the compactification
scale $\frac1R$, leading to the following hierarchy
\begin{equation}
 M_{\rm GUT} \le\textstyle\frac1R\le\frac1r<M_s, M_P
\end{equation}
where we have not specified the hierarchy between $M_s$ and $M_P$.

In Sec. \ref{sec:Sflavor}, we discuss phenomenological requirements
in the GMSB scenario toward the SUSY flavor problem. In Sec.
\ref{sec:Model}, we present a  $\Z_{12-I}$  example. In Sec.
\ref{sec:SU(5)}, we discuss the hidden sector gauge group SU(5)$'$
where a GMSB spectrum is possible.

\section{SUSY FCNC conditions and gauge mediation}\label{sec:Sflavor}

The MSSM spectrum between the SUSY breaking and GUT scales fixes the
unification coupling constant $\alpha_{\rm GUT}$ of the observable
sector at around $\frac{1}{25}$. If a complete SU(5) multiplet in
the observable sector is added, the unification is still achieved
but the unification coupling constant will become larger. Here, we
choose the unification coupling constant in the range $\alpha_{\rm
GUT}\sim \frac{1}{30}-\frac{1}{20}$.

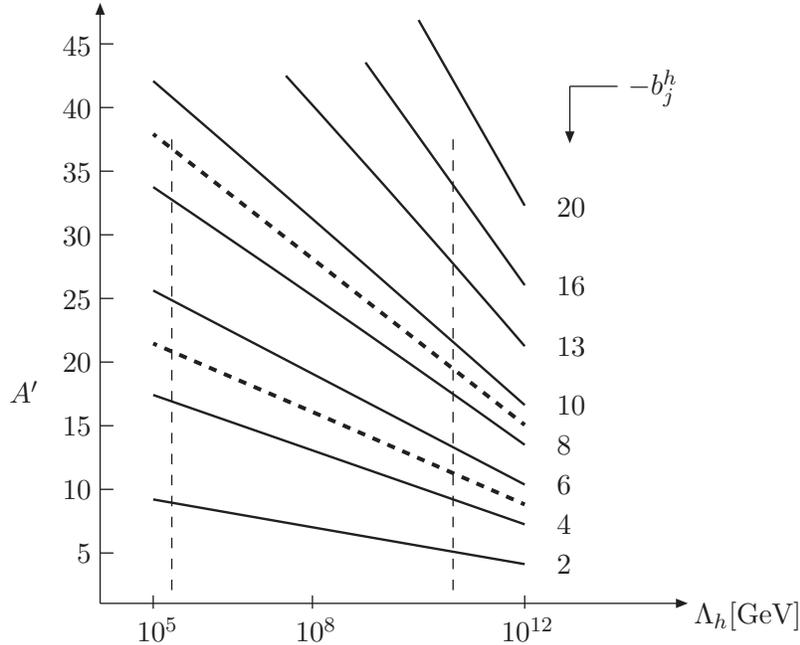
\begin{figure}[t]
\begin{center}
\begin{picture}(400,230)(0,0)

\LongArrow(80,5)(80,230)\Text(57,85)[r]{$A'$}
\LongArrow(80,5)(300,5)\Text(305,0)[l]{$\Lambda_h$[GeV]}
\Line(100,3)(100,8)\Text(102,-5)[c]{$10^5$}
\Line(160,3)(160,8)\Text(162,-5)[c]{$10^8$}\Line(240,3)(240,8)
\Text(242,-5)[c]{$10^{12}$}

\Line(80,24)(85,24)\Text(77,24)[r]{5}
\Line(80,48)(85,48)\Text(77,48)[r]{10}
\Line(80,72)(85,72)\Text(77,72)[r]{15}
 \Line(80,96)(85,96)\Text(77,96)[r]{20}
\Line(80,120)(85,120)\Text(77,120)[r]{25}
\Line(80,144)(85,144)\Text(77,144)[r]{30}
\Line(80,168)(85,168)\Text(77,168)[r]{35}
\Line(80,192)(85,192)\Text(77,192)[r]{40}
\Line(80,216)(85,216)\Text(77,216)[r]{45}

 \Text(280,200)[l]{$-b_j^h$}
\Line(275,200)(257,200)\LongArrow(257,200)(257,180)
 {\SetWidth{0.9}
 \Curve{(100,44.2)(160,33.7)(240,19.8)} \Text(253,19.8)[l]{$2$}
 \Curve{(100,83.6)(160,62.7)(240,34.8)} \Text(253,34.8)[l]{$4$}
{\SetWidth{1.5}\DashCurve{(100,103)(160,77.2)(240,42.3)}{3} }
\Curve{(100,123)(160,91.6)(240,49.8)} \Text(253,50)[l]{$6$}
\Curve{(100,162)(160,121)(240,64.8)}  \Text(253,65)[l]{$8$}
{\SetWidth{1.5}\DashCurve{(100,182)(160,135)(240,72.3)}{3} }
\Curve{(100,202)(160,150)(240,79.8)} \Text(253,80)[l]{$10$}
\Curve{(150,204)(200,148)(240,102)} \Text(253,102)[l]{$13$}
\Curve{(180,209)(200,181)(240,125)} \Text(253,125)[l]{$16$}
\Curve{(200,225)(220,190)(240,155)} \Text(253,155)[l]{$20$} }

\DashLine(107,10)(107,180){4} \DashLine(213,10)(213,180){4}

\end{picture}
\caption{Constraints on $A'$. The confining scale is defined as the
scale $\mu$ where $\alpha^h_j(\mu)=1$. Using $\xi=0.1, M_X=2\times
10^{16}$ GeV in the upper bound region and $\xi=0.1,
M_X=\frac12\times 10^{6}$ GeV in the lower bound region, we obtain
the region bounded by dashed vertical lines. Thick dash curves are
for $-b_j^h=5$ and 9. }\label{fig:hcoupl}
\end{center}
\end{figure}

The GMSB scenario has been adopted to hide the gravity mediation
below the GMSB effects so that SUSY breaking need not introduce
large flavor changing neutral currents (FCNC) \cite{DineNeslson}:
\begin{align}
&\frac{\Lambda_{h}^3}{M_P^2}\le 10^{-3}\ {\rm TeV}\Rightarrow
\Lambda_{h}\le
2\times 10^{12}\ {\rm GeV} \\
&\frac{(\xi \Lambda_{h})^2}{M_X}\sim 10^{3}\ {\rm GeV}
\end{align}
where $M_P$ is the reduced Planck mass $2.44\times 10^{18}$ GeV,
$M_X$ is the effective messenger scale (including coupling
constants) in the GMSB scenario, $M_X\ge \frac12\times 10^{6}$ GeV
for acceptable FCNC effects, and $\xi$ measures the hidden sector
squark condensation scale compared to the hidden sector confining
scale. So, a possible range of $\Lambda_h$ is $\Lambda_h$ =
[$0.7\times 10^5\xi^{-1}$ GeV, $2\times 10^{12}$ GeV]. Because of
the SUSY breaking scale fixed at TeV, the messenger scale $M_X$ is a
function of $\Lambda_h$. These conditions on the confining scale of
the hidden sector fix the strength of the hidden sector unification
coupling constant $\alpha_{\rm GUT}^{h}$. The GUT scale coupling
constant is related to the coupling at scale $\mu$, at one loop
order, by
\begin{align}
\frac{1}{\alpha_{\rm
GUT}^{h}}=\frac{1}{\alpha_j^{h}(\mu)}+\frac{-b_j^{h}}{2\pi}
\ln\left|\frac{M_{\rm GUT}^{h}}{\mu} \right|. \label{hcoupl}
\end{align}
Now the expression (\ref{hcoupl}) is used to give constraint on
$\alpha_{\rm GUT}^{h}$. Defining the inverse of unification coupling
constants as
\begin{equation}
A=\frac{1}{\alpha_{\rm GUT}},\ A'=\frac{1}{\alpha_{\rm GUT}^{h}},
\end{equation}
we express $A'$ in terms of the scale $\Lambda_h$ as\footnote{One
can determine $\Lambda_h$ where $\alpha_h=\infty$ for which near
$\Lambda_h$ the one loop estimation is not valid. So we estimate
$\Lambda_h$ at $\alpha_h=1$.}
\begin{align}
A'-1=\frac{-b_j^h}{2\pi}\ln\left(\frac{M_{\rm GUT}^{h}}{\Lambda_h}
 \right).
\end{align}
If $M_{\rm GUT}\simeq 2\times 10^{16}$ GeV and $\Lambda_h\simeq
2\times 10^{10}$ GeV, we obtain $A'$ in terms of $-b_j^h$ as shown
in  Eq. (\ref{Aprime}).
\begin{equation}
\begin{array}{cccccccccc}
-b_j^h& A' &\quad -b_j^h& A'  &\quad -b_j^h& A' &\quad -b_j^h& A'
 &\quad -b_j^h& A'\\
  2& 5.4 &\quad 4 & 9.8 &\quad 6 & 14.2 &\quad 8 & 18.6
   &\quad 10 & 23.0\\
     12 & 27.4 &\quad 14 & 31.8 &\quad 16 & 36.2
  &\quad 18& 40.6&\quad  20 & 45.0
\end{array}\label{Aprime}
\end{equation}
In Fig. \ref{fig:hcoupl} we present figures  of $A'$ versus
$\Lambda_h$ for several values of $-b_j^h$.

The GMSB relies on dynamical supersymmetry breaking (DSB)
\cite{Shirman}. The well-known DSB models are an SO(10)$'$ model
with ${\bf 16}'$ or ${\bf 16}'+{\bf 10}'$, and an SU(5)$'$ model
with ${\bf 10}'+{\bf\overline{5}}'$. If we consider a metastable
vacuum, a SUSY QCD type is possible in SU(5)$'$ with six or seven
flavors, $6(\bf 5'+\overline{5}')$ or $7(\bf 5'+\overline{5}')$
\cite{ISS}. The reason that we have this narrow band of $N_f$ is
that the theory must be infrared free in a controllable way in the
magnetic phase. Three family models with $\alpha'<\frac1{25}$ are
very rare, and we may allow at most up to 20\% deviation from
$\alpha_{\rm GUT}$ value, i.e. $\alpha'>\frac1{30}$. Then, from Fig.
\ref{fig:hcoupl} we note that it is almost impossible to have an
SO(10)$'$ model from superstring toward the GMSB. The reason is that
SO(10)$'$ matter representations from superstring are not big and
hence $-b_j=24-\sum_i l(R_i)$ seems very large. The flipped SU(5)
model of Ref. \cite{KimKyaegut} has one ${\bf 16}'$ and one ${\bf
10}'$ of SO(10)$'$ with $-b^h_{\rm SO(10)}=21$, which can lead to a
GMSB if the hidden sector coupling at the GUT scale is very small,
$\alpha_{\rm GUT}^{h}<\frac{1}{33}$. On the other hand, SU(5) models
can have many possibilities with $-b^h_{\rm SU(5)}=15-N_f$. The
SU(5) model with seven flavors gives $-b^h_{\rm SU(5)}=8$, which
allows a wide range of $\Lambda_h$. It is even possible to have
$\alpha_{\rm GUT}^{h}=\alpha_{\rm GUT}\simeq \frac{1}{25}$ for
$\Lambda_h\sim 3\times 10^7$ GeV with the messenger scale $M_X$
around $10^{12}$ GeV. Bigger SU($N$)$'$ groups with $N>5$ are also
possible for the ISS scenario, but it is difficult to obtain many
flavors of SU(N)$'$ in orbifold compactification. Most orbifold
models have chiral fields at the order of 200 fields (among which
many are singlets) and if we go to large SU($N$)$'$ groups it is
more difficult to obtain a large number of SU($N$)$'$ flavors with
the required three families of quarks and leptons.

The ISS type models are possible for  SO($N_c$) and Sp($N_c$) groups
also \cite{ISS}. In this paper, however we restrict our study to the
SU(5)$'$ hidden sector only. We just point out that SO($N_c$)
groups, with the infrared free condition in the magnetic phase for
$N_f<\frac32(N_c-2)$, are also very interesting toward the unstable
vacua, but the study of the phase structure here is more involved.
On the other hand, we do not obtain Sp($N_c$) groups from orbifold
compactification of the hidden sector \Eh.

\section{A $\Z_{12-I}$ Model}\label{sec:Model}

We illustrate an SSM from ${\bf Z}_{12-I}$. The twist vector in the
six dimensional (6d) internal space is
\begin{align}
{\rm \Z_{12-I}\ shift}:\quad
\phi&=\textstyle(\frac{5}{12}~\frac{4}{12}~\frac{1}{12}).
\end{align}
The compactification radius of (12)- and (56)-tori is $ r$ and the
compactification radius of (34)-torus is $R$, with a hierarchy of
radii $r\ll R$.

 We obtain the 4D gauge group by considering massless
conditions satisfying $P\cdot V=0$ and $P\cdot a_3=0$ in the
untwisted sector \cite{DHVW:1985}. This gauge group is also obtained
by considering the common {\it intersection} of gauge groups
obtained at each fixed point.

We embed the discrete action $\Z_{12-I}$ in the \EE\ space in terms
of the shift vector $V$ and the Wilson line $a_3$
as\footnote{Another interesting standard model from $\Z_{12-I}$ can
be found in \cite{KimKyaeSM}.}
\begin{align}
 V&=\textstyle
 \frac1{12}(2~2~2~4~4~1~3~6)(3~3~3~3~3~1~1~1)'
 \label{Z12ImodelC}\\
  a_3&=\textstyle
\frac{1}{3}(0~0~0~0~0~0~0~0)(
 0~0~0~0~0~2~{\textstyle -1}~{\textstyle -1})'.\label{WilsonLine}
\end{align}

\noindent {\bf  (a) Gauge group}:
 The 4D gauge groups are obtained by $P^2=2$ vectors
satisfying $P\cdot V=0$ and $P\cdot a_3=0$ mod integer,
\begin{align}
SU(4)\times SU(2)_W &\times SU(2)_V\times SU(2)_n\times U(1)_a
\times
U(1)_b\nonumber \\
&\times [SU(5)\times SU(3)\times U(1)^2]'.
\end{align}
The simple roots of SU(4), SU(2)$_W$, SU(2)$_V$, and SU(2)$_n$
are\footnote{We will use the representations $\four,\fourb$ and
$\six$ of SU(4) as the complex conjugated ones obtained from Eq.
(\ref{SU4roots}) but still keep the U(1) charges so that $t,b,e,$
etc. are shown instead of $t^c, b^c, e^c,$ etc.}
\begin{align}
&{SU(4):}\left\{
\begin{array}{cc}
\alpha_1=& (0\ 1\ {\textstyle -1}\ 0\ 0\ ;\ 0\ 0\ 0)\\
\alpha_2=& (\frac{1}{2}\ \frac{-1}{2}\ \frac{1}{2}\ \frac{1}{2}\
\frac{1}{2}\ ;\  \frac{-1}{2}\ \frac{-1}{2}\ \frac{-1}{2})\\
\alpha_3=&  (\frac{1}{2}\ \frac{-1}{2}\ \frac{-1}{2}\ \frac{-1}{2}\
\frac{-1}{2}\ ;\  \frac{1}{2}\ \frac{1}{2}\ \frac{1}{2})\\
\end{array}\right. \label{SU4roots}\\
&{SU(2)_W:}
\begin{array}{cc}
\alpha_W=& (0~0~0~1~ {\textstyle -1};~0~0~0)
\end{array}\\
&{SU(2)_V:}
\begin{array}{cc}
\alpha_V=& (\frac{1}{2}\ \frac{1}{2}\ \frac{1}{2}\ \frac{1}{2}\
\frac{1}{2}\ ;\  \frac{1}{2}\ \frac{1}{2}\ \frac{1}{2})
\end{array}
\\
&{SU(2)_n:}
\begin{array}{cc}
\alpha_n=& (\frac{1}{2}\ \frac{1}{2}\ \frac{1}{2}\ \frac{-1}{2}\
\frac{-1}{2}\ ;\  \frac{-1}{2}\ \frac{-1}{2}\ \frac{1}{2})
\end{array}.
\end{align}
The SU(2)$_V$ is like SU(2)$_R$ in the Pati-Salam(PS) model
\cite{PatiSalam}. The gauge group SU(4) will be broken by the vacuum
expectation value (VEV) of the neutral singlet in  the PS model. In
the PS model, the hypercharge direction is
\begin{equation}
Y=\tau_3+Y_4+Y'
\end{equation}
where $\tau_3$ is the third SU(2)$_V$ generator, $Y_4$ is an SU(4)
generator, e.g. for $\four$,
\begin{equation}
Y_4=\textstyle{\rm diag.}(\frac16~\frac16~\frac16~\frac{-1}{2}),
\end{equation}
and $Y'$ is a hidden-sector E$_8'$ generator. We find that exotics
cannot be made vectorlike if we do not include $Y'$. We succeed in
making the model exotics-free by choosing $Y'$ as
\begin{equation}
Y'=\textstyle
(0^8)(\frac{1}{3}~\frac{1}{3}~\frac{1}{3}~\frac{1}{3}~\frac{1}{3}
~0^3)'.
\end{equation}
Note that SU(2)$_V$ doublet components have the unit hypercharge
difference. Two U(1) charges of \Eo\ are  obtained by taking scalar
products with
\begin{align}
&Q_a\to\textstyle
(0~0~0~0~0~1~{\textstyle -1}~0)\\
& Q_b\to\textstyle ({\textstyle 1}~{\textstyle 1}~{\textstyle
1}~{\textstyle -1}~{\textstyle -1}~1~1~{\textstyle -3}).
\end{align}

\noindent {\bf  (b) Matter representations}:
 Now there is a standard
method to obtain the massless spectrum in $\Z_{12-I}$ orbifold
models. The spectra in the untwisted sectors $U_1, U_2,$ and $U_3$,
and twisted sectors, $T1_{0,+,-},T2_{0,+,-},T3, T4_{0,+,-},
T5_{0,+,-},$ and $T6$, are easily obtained \cite{KimKyaeSM}. The
representations are denoted as
\begin{equation} [{\bf SU(4)},{\bf SU(2)}_W,{\bf
SU(2)}_V; SU(2)_n;{\bf SU(5)}',SU(3)'],
\end{equation}
and for obvious cases we use the standard PS notation
\begin{equation}
({\bf SU(4)},{\bf SU(2)}_W,{\bf SU(2)}_V)_{Y'}.
\end{equation}
We list all matter fields below,
\begin{align}
\begin{array}{l}
U_1:~ (\fourb,\two,\one)_0,\ 2(\six,\one,\one)_0\\
 U_2:~ 2(\four,\one,\two)_0,\ (\six,\one,\one)_0\\
 U_3:~ (\four,\one,\two)_0,\ 2(\one,\two,\two)_0,\ (\one,\one,\one;\two;
 \one,\one)_0\\
 T_{1_0}:~ (\fourb,\one,\one)_{1/2},\ (\one,\two,\one)_{1/2},
 \ (\one,\one,\two)_{1/2}\\
 T_{1_+}:~  (\one,\two,\one)_{-1/2},
 \ (\one,\one,\two)_{-1/2}\\
 T_{1_-}:~ (\one,\one,\two;\one;\five';\one)_{-1/10}\\
 T_{2_0}:~  (\six,\one,\one)_{0},\
\two^n_0,\ \one_0\\
 T_{2_+}:~ \five'_{2/5},\ \threeb'_0,\ \quad  \\
T_{2_-}:~ (\one,\two,\two)_0,\ \three'_0,\ \two^n_0,\ 2\cdot\one_0\\
 T_3:~  (\fourb,\one,\one)_{1/2},\ (\four,\one,\one)_{-1/2},\
  (\four,\one,\one)_{1/2},\ 2(\fourb,\one,\one)_{-1/2},\
3(\one,\two,\one)_{1/2},\\
\quad\quad  2(\one,\two,\one)_{-1/2},\
2(\one,\one,\two;\two;\one;\one)_{1/2},\
   (\one,\one,\two;\two;\one;\one)_{-1/2},\\
\quad\quad  (\one,\two,\one;\one;\five';\one)_{-1/10},\
 2\cdot(\one,\two,\one;\one;\fiveb';\one)_{1/10}  \\
T_{4_0}:~ 2(\one,\one,\one;\two;\one;\threeb')_0,\ 2\cdot\threeb'_0 \\
T_{4_+}:~ 2(\fourb,\two,\one)_0,\ 2(\four,\one,\two)_0,\
2(\six,\one,\one)_0,\ 7\cdot\two^n_0,
\ 9\cdot\one_0\\
T_{4_-}:~ 2(\one,\one,\one;\two;\one;\three')_0,\ 2\cdot\three'_0\\
T_{7_+}:~ (\fourb,\one,\one)_{1/2},\ (\one,\one,\two)_{1/2}\\
T_{7_-}:~ (\fourb,\one,\one)_{-1/2},\
(\one,\one,\two;\two;\one;\one)_{-1/2},\ (\one,\one,\two)_{-1/2} \\
T_6:~ 6\cdot\fiveb'_{-2/5},\ 5\cdot\five'_{2/5},\
\\
 \end{array}\label{Allspectrum}
 \end{align}
where $\one=(\one,\one,\one;\one;\one;\one),\two^n=
(\one,\one,\one;\two;\one;\one),
\three'=(\one,\one,\one;\one;\one;\three')$ and
$\threeb'=(\one,\one,\one;\one;\one;\threeb')$.
 \begin{table}[t]
\begin{center}
\begin{tabular}{|c|c|c|c|c|}
\hline  $P+[4V+4a]$  & $\chi$ & No.$\times$(Repts.)$_{Y,Q_1,Q_2}$
& PS rep. &Label\\
\hline
 $(\underline{\frac12~\frac{-1}{2}~\frac{-1}{2}}~
 \underline{\frac{1}{2}~\frac{-1}{2}}~\frac12
 ~\frac{-1}{2}~\frac12)_{U_1}$
  & $L$ & $
 (\overline{\bf 3},{\bf 2},\one;1;{\bf 1}, 1)_{-1/6,1,-2}^L$
 & $(\fourb,\two,\one)_0$ &$\bar q_3$
\\
$(0~0~0~\underline{1~0}~0~0~{\textstyle -1})_{U_1}$  & $L$ &$
 ({\bf 1},{\bf 2},\one;1;{\bf 1}, 1)_{1/2,0,4}^L$
 & $(\fourb,\two,\one)_0$ &$\bar l_3$
\\
 $(\underline{\frac12~\frac{1}{2}~\frac{-1}{2}}~
 {\frac{1}{2}~\frac{1}{2}}~\frac12
 ~\frac{1}{2}~\frac{-1}{2})_{U_2}$
  & $L$ & $
 ({\bf 3},{\bf 1},\uparrow;1;{\bf 1}, 1)_{2/3,0,2}^L$
 & $(\four,\one,\two)_0$ &$t$
\\
$(\underline{{\textstyle -1}~0~0}~0~0~0~0~{\textstyle -1})_{U_2}$ &
 $L$ &$
 ({\bf 3},{\bf 1},\downarrow;1;{\bf 1}, 1)_{-1/3,0,2}^L$
 & $(\four,\one,\two)_0$ &$b$
\\
$(\frac{-1}{2}~\frac{-1}{2}~\frac{-1}{2}~
\frac{-1}{2}~\frac{-1}{2}~\frac{1}{2}~\frac{1}{2}~\frac{-1}{2})_{U_2}$
&
 $L$ &$
 (\one,\one,\downarrow;1;{\bf 1}, 1)_{-1,0,2}^L$
 & $(\four,\one,\two)_0$ &$\tau$
\\
$(0~0~0~0~0~1~1~0)_{U_2}$ &
 $L$ &$
 (\one,\one,\uparrow;1;{\bf 1}, 1)_{0,0,2}^L$
 & $(\four,\one,\two)_0$ &$\nu_0$
\\
$(\underline{0~1~1}~ 0~0~0~0~0)_{U_2}$
  & $L$ & $
 ({\bf 3},{\bf 1},\uparrow;1;{\bf 1}, 1)_{2/3,0,2}^L$
 & $(\four,\one,\two)_0$ &$(c)$
\\
$(\underline{\frac{-1}{2}~\frac{1}{2}~\frac{1}{2}}~\frac{-1}{2}~
\frac{-1}{2}~\frac{-1}{2}~\frac{-1}{2}~\frac{-1}{2})_{U_2}$  & $L$
&$
 ({\bf 3},{\bf 1},{\downarrow};1;{\bf 1}, 1)_{-1/3,0,2}^L$
 & $(\four,\one,\two)_0$ &$(s)$
\\
$(0~0~0~{\textstyle -1}~{\textstyle -1}~0~0~0)_{U_2}$ &
 $L$ &$
 (\one,\one,\downarrow;1;{\bf 1}, 1)_{-1,0,2}^L$
 & $(\four,\one,\two)_0$ &$(\mu)$
\\
$(\frac{1}{2}~\frac{1}{2}~\frac{1}{2}~\frac{-1}{2}~\frac{-1}{2}~
\frac{1}{2}~\frac{1}{2}~\frac{1}{2})_{U_2}$ &
 $L$ &$
 (\one,\one,\uparrow;1;{\bf 1}, 1)_{0,0,2}^L$
 & $(\four,\one,\two)_0$ &$\nu_0$
\\
$(0~0~0~\underline{1~0}~0~{\textstyle 1}~0)_{U_3}$  & $L$ &$
 ({\bf 1},{\bf 2},{\uparrow};1;{\bf 1}, 1)_{1/2,-1,0}^L$
 & $(\one,\two,\two)_0$ &$H_u$
\\
$(0~0~0~
 \underline{0~{\textstyle -1}}~0
 ~{\textstyle -1}~0)_{U_3}$
 &  $L$ & $
 ({\bf 1},{\bf 2},{\bf \downarrow};1;{\bf 1}, 1)_{-1/2,-1,0}^L$
 & $(\one,\two,\two)_0$ &$H_d$
\\[0.2em]
\hline $(\underline{\frac23~\frac{-1}{3}~\frac{-1}{3}}~
 \underline{\frac{1}{3}~\frac{-2}{3}}~0~0~0)_{T4_+}$
 &  $L$ & $
 2(\overline{\bf 3},{\bf 2},\one;1;{\bf 1}, 1)_{-1/6,0,1/3}^L$
 & $(\fourb,\two,\one)_0$ &$\bar q_2,\ \bar q_1$
\\
 $(\underline{\frac23~\frac{2}{3}~\frac{-1}{3}}~
 \frac{1}{3}~\frac{1}{3}~\frac{1}{3}~0~0)_{T4_+}$
 &  $L$ & $
 2({\bf 3},{\bf 1},{\uparrow};1;{\bf 1}, 1)_{2/3,1/3,2/3}^L$
 & $(\four,\one,\two)_0$ &$(c),\ u$
\\ $(\underline{\frac{1}{6}~\frac{1}{6}~\frac{-5}{6}}~
\frac{-1}{6}~\frac{-1}{6}~\frac{-1}{6}~\frac{-1}{2}~
\frac{-1}{2})_{T4_+}$
 &  $L$ & $
 2({\bf 3},{\bf 1},{\downarrow};1;{\bf 1}, 1)_{-1/3,1/3,2/3}^L$
 & $(\four,\one,\two)_0$ &$(s), \ d$
\\
 $(\frac16~\frac16~\frac16~
 \underline{\frac{5}{6}~\frac{-1}{6}}~\frac{-1}{6}~\frac{-1}{2}~
 \frac{-1}{2})_{T4_+}$
 & $L$ & $
 2({\bf 1},{\bf 2},{\bf 1};1;{\bf 1}, 1)_{1/2,1/3,2/3}^L$
 & $(\fourb,\two,\one)$ &$\bar l_2,\ \bar l_1$
\\
 $(\frac{-1}{3}~\frac{-1}{3}~\frac{-1}{3}~\frac{-2}{3}~\frac{-2}{3}~
 \frac{1}{3}~0~0)_{T4_+}$
 &  $L$ & $
 2({\bf 1},{\bf 1},\downarrow;1;{\bf 1}, 1)_{-1,1/3,2/3}^L$
 & $(\four,\one,\two)_0$ &$(\mu),\ e$
\\
 $(\frac{1}{6}~\frac{1}{6}~\frac{1}{6}~\frac{-1}{6}~
 \frac{-1}{6}~\frac{5}{6}~
 \frac{1}{2}~\frac{1}{2})_{T4_+}$
 &  $L$ & $
 2({\bf 1},{\bf 1},\uparrow;1;{\bf 1}, 1)_{0,1/3,2/3}^L$
 & $(\four,\one,\two)_0$ &$2\nu_0$
\\[0.2em]
\hline
\end{tabular}
\end{center}
\caption{Some conventionally charged massless states in $U$ and
$T4_+$. Out of four $Q_{\rm em}=\frac23$ quarks (and $-\frac13$
quarks and --1 leptons) of this table, only three combinations form
families, i.e. one combination from bracketed ones. The VEVs of
$\nu_0$s break SU(4) down to SU(3)$_c$.} \label{table:Families}
\end{table}
In the model, there does not appear any exotics.\footnote{We found
another exotics free model by including $Y'$ in the hypercharge $Y$
\cite{KimKyaeSM}.} All SU(5)$'$ singlet fields carry the standard
charges, i.e. quarks with \Qem=$\frac23,-\frac13$ and leptons and
Higgs with \Qem$=0,\pm1$. The real representation $\six$ of SU(4)
carries \Qem$=-\frac13$ for $\three$ and \Qem$=\frac13$ for
$\threeb$. Thus, this model is exotics free. The classification of
the particles is along Pati-Salam, but it is not the Pati-Salam
model \cite{PatiSalam} since it is not symmetric under
SU(2)$_W\leftrightarrow{\rm SU(2)}_V$. In addition, the hypercharge
$Y'$ belongs to \Eh\ and hence
SU(4)$\times$SU(2)$_W\times$SU(2)$_V\times$U(1)$_{Y'}$ cannot belong
to an SO(10). The SU(5)$'$ singlet fields do not have any
SU(3)$_c\times$SU(2)$_W\times$U(1)$_Y$ gauge anomaly. For example,
six lepton doublets $\overline{l}_{1/2}$ from $U_1, T_3$ and
$T_{4_+}$ and three anti-doublets $l_{-1/2}$ from $T_{1_+}$ and
$T_3$, lead to lepton doublets of three families. The charge $\pm 1$
leptons ($e^\pm$) appear as twelve $e^-$ from $2U_2, 1U_3, 1T_{1_+},
3T_3, 2T_{4_+}, 3T_{5_-}$ and nine $e^+$ from $2T_{1_0},5T_3,
2T_{5_+}$, and three $e^-$s are left. Thus, these leptons do not
have the SM gauge anomaly. If composite leptons are made from
$\five'$ and $\fiveb'$, they must be anomaly free by themselves.

As shown in Table \ref{table:Families}, the model has three families
of the SSM, one in the untwisted sector and two in the twisted
sector. Breaking of SU(4) down to SU(3)$_c$ is achieved by VEVs of
neutral components in $(\four,\one,\one)_{1/2}\equiv V_1,
(\four,\one,\two)_0\equiv{V}_2, (\fourb,\one,\one)_{-1/2}\equiv
\overline{V}_1,
 (\one,\one,\two)_{1/2}\equiv v$
and $(\one,\one,\two)_{-1/2}\equiv \overline{v}$. A SUSY $D$-flat
direction at the GUT scale requires $V_1^2+{V}_2^2=\overline{V}_1^2,
v^2={V}_2^2+\overline{v}^2$, and $V_1^2+v^2=\overline{V}_1^2+\bar
v^2$. Certainly, these conditions can be satisfied. At this point,
we are content merely with having three SSM families without
exotics, and let us proceed to discuss SUSY breaking via the GMSB
scenario, using the hidden sector SU(5)$'$.

\section{Hidden sector SU(5)$'$}\label{sec:SU(5)}

 As shown in Table
\ref{table:Hidden}, there are ten ${\bf 5}'$s and ten $\overline{\bf
5}'$s. But some of these obtain masses by Yukawa couplings. The
H-momenta of the fields from the sectors are
\cite{Katsuki,KimKyaeSM,ChoiKimBk}
\begin{align}
&U_1: (-1,0,0),\quad U_2: (0,1,0),\quad U_3:
(0,0,1),\nonumber\\
&\textstyle T_1:(\frac{-7}{12},\frac{4}{12},\frac1{12}),\quad
 T_2:(\frac{-1}{6},\frac46,\frac16),\quad T_3:
 (\frac{-3}{4},0,\frac{1}{4}),\nonumber\\
&\textstyle
 T_4:(\frac{-1}{3},\frac13,\frac13),\quad
\left\{T_5:(\frac{1}{12},\frac{-4}{12},\frac{-7}{12})\right\}, \quad
T_6:(\frac{-1}{2},0,\frac12),\\
&\textstyle T_7:(\frac{-1}{12},\frac{4}{12},\frac{7}{12}),\quad
T_9:(\frac{-1}{4},0,\frac{3}{4}) , \nonumber
\end{align}

Therefore, from the H-momentum rule alone, the cubic Yukawa
couplings $ T_3T_9U_2$ and $T_6T_6U_2$ are expected for ${\bf 5}'$s
and $\overline{\bf 5}'$s appearing in $T_3, T_9,$ and $T_6$, if they
make the total H-momentum $(-1, 1, 1)$ mod
$(12,3,12)$.\footnote{Details of the rules for $\Z_{12-I}$ are given
in \cite{KimKyaegut,KimKyaeSM}.}
 However, the gauge symmetry
forbids them at the cubic level. But we expect that the Yukawa
couplings appear at higher orders. For example, to make $H=(-1,1,1)$
we can multiply $T_3T_9$ or $T_6T_6$ times
\begin{equation}
(\four,\one,\two)_0^{(U_2)}(\fourb,\one,\one)_{-1/2}^{(T_{7_-})}
(\one,\one,\two)_{1/2}^{(T_{1_0})}T_{4_+}(T_{4_0}T_{4_0}T_{4_0})^{11}
\label{Yukex}
\end{equation}
where $T_{4_+}$ is $\one_0$ and $T_{4_0}$ is $\threeb_0$ and
$T_{4_0}T_{4_0}T_{4_0}=\epsilon^{\alpha\beta\gamma}\threeb_{0\alpha}
\threeb_{0\beta}\threeb_{0\gamma}$. Every field in the above has
neutral components which can develop a large VEV.
 \begin{table}[t]
\begin{center}
\begin{tabular}{|c|c|c|}
\hline  $P+n[V\pm a]$ &  $\chi$ & No.$\times$(Repts.)$_{Y,Q_1,Q_2}$
\\
\hline $(\underline{\frac16~\frac16~\frac16~\frac{1}{3}~\frac{1}{3}~
 \frac{1}{12}~\frac{1}{4}
 ~\frac{1}{2}})
 (\underline{\frac34~\frac{-1}{4}~\frac{-1}{4}~\frac{-1}{4}
 ~\frac{-1}{4}}~
 \frac{-1}{4}~\frac{-1}{4}~\frac{-1}{4})'_{T1_-}$
 & $L$ & $
 ({\bf 1},{\bf 1},\two;1;{\bf 5}', 1)_{-1/10,-1/6,-4/3}^L$
\\
$(\frac{-1}{6}~\frac{-1}{6}~\frac{-1}{6}~~\frac{1}{6}~\frac{1}{6}~
 \frac{-1}{3}~0~\frac{1}{2} )
 (\underline{1~0~0~0 ~0}~
 0~0~0)'_{T2_+}$
  & $L$ & $({\bf 1},{\bf 1},{\bf 1};1;{\bf 5}', 1)_{2/5,-1/3,-8/3}^L$
\\
 $(0~0~0~\underline{\frac12~\frac{-1}{2}}~\frac{-1}{4}~\frac{1}{4}~0)
 (\underline{\frac34~\frac{-1}{4}~\frac{-1}{4}~\frac{-1}{4}
 ~\frac{-1}{4}}~\frac{1}{4}~\frac{1}{4}
 ~\frac{1}{4})'_{T3}$
  & $L$ & $({\bf 1},{\bf 2},{\bf 1};1;{\bf 5}', 1)_{-1/10,-1/2,0}^L$
\\
 $(0~0~0~\underline{\frac12~\frac{-1}{2}}~\frac{1}{4}~\frac{-1}{4}~0)
 (\underline{\frac{-3}{4}~\frac{1}{4}~\frac{1}{4}~\frac{1}{4}
 ~\frac{1}{4}}~\frac{-1}{4}~\frac{-1}{4}
 ~\frac{-1}{4})'_{T9}$ & $L$ & $
 2({\bf 1},{\bf 2},{\bf 1};1;\overline{\bf 5}', 1)_{1/10,1/2,0}^L$
\\
$(0~0~0~0~0~\frac{-1}{2}~\frac{1}{2}~0)
 (\underline{\textstyle -1~0~0~0~0}~0~0~0)'_{T6}$ & $L$ & $
 4({\bf 1},{\bf 1},{\bf 1};1;\overline{\bf 5}', 1)_{-2/5,-1,0}^L$
\\
$(0~0~0~0~0~\frac{-1}{2}~\frac{1}{2}~0)
 (\underline{\textstyle 1~0~0~0~0}~0~0~0)'_{T6}$ & $L$ & $
 2({\bf 1},{\bf 1},{\bf 1};1;{\bf 5}', 1)_{2/5,-1,0}^L$
\\
$(0~0~0~0~0~\frac{1}{2}~\frac{-1}{2}~0)
 (\underline{\textstyle -1~0~0~0~0}~0~0~0)'_{T6}$ & $L$ & $
 2({\bf 1},{\bf 1},{\bf 1};1;\overline{\bf 5}', 1)_{-2/5,1,0}^L$
\\
$(0~0~0~0~0~\frac{1}{2}~\frac{-1}{2}~0)
 (\underline{\textstyle 1~0~0~0~0}~0~0~0)'_{T6}$ & $L$ & $
 3({\bf 1},{\bf 1},{\bf 1};1;{\bf 5}', 1)_{2/5,1,0}^L$
\\[0.2em]
\hline
\end{tabular}
\end{center}
\caption{Hidden sector SU(5)$'$ representations. We picked up the
left-handed chirality only from $T_1$ to $T_{11}$ representations.}
\label{table:Hidden}
\end{table}

Out of ten SU(5)$'$ quarks, there may result any number of very
light ones according to the choice of the vacuum. A complete study
is very complicated and here we just mention that it is possible to
have six or seven light SU(5)$'$ quarks out of ten. The point is
that we have enough SU(5)$'$ quarks. For example, one may choose the
$T_3T_9$ coupling such that one pair of SU(2)$_W$ doublets (two
SU(5)$'$ quarks) becomes heavy with a mass scale of $m_1$. For the
sake of a concrete discussion, presumably by fine-tuning at the
moment, one may consider the $T_6T_6$ coupling such that the
following $ {\bf 5}'\cdot \overline{\bf 5}'$ mass matrix form
\begin{align}
\left(
\begin{array}{cccccc}
m_1&m_1&0&0&0&0 \\ m_1&m_1&0&0&0&0 \\
0&0&m_2&m_2&m_2&m_3\\
0&0&m_2&m_2&m_2&m_3\\
0&0&m_2&m_2&m_2&m_3
\end{array}
\right)\label{massex}
\end{align}
where 0 entries are due to the U(1)$_a$ charge consideration. If so,
out of five ${\bf 5}'$s and six $\overline{\bf 5}'$s from $T_6$
three ${\bf 5}'$s and four $\overline{\bf 5}'$s remain massless, one
pair of ${\bf 5}'$ and $\overline{\bf 5}'$ obtain mass $2m_1$ and
another pair obtain mass $3m_2$ if $m_3=0$. Thus, the mass pattern
of the total ten flavors of SU(5)$'$ hidden sector quarks of Table
\ref{table:Hidden} will be six light SU(5)$'$ quarks and four
massive SU(5)$'$ quarks. Choosing a different vacuum, another set of
massless SU(5)$'$ quarks would be obtained. In this consideration,
the location of fields at fixed points and the permutation
symmetries  must be considered. For example,  the $T_6$ sector being
basically $\Z_2$ in the (12)- and (56)-tori has four fixed points in
the (12)- and (56)-tori. These may be classified by the permutation
symmetry $S_4$ \cite{permsym}. The $S_4$ representations are
$\one,\one',\two,\three$ and $\three'$. The four fixed points can be
split into $\three+\one$ or to $\two+\one+\one'$. The combination of
(12)- and (56)-tori can have $\three\otimes\three=\three \oplus
\three' \oplus\two \oplus\one$. Thus, the $T_6$ sectors can contain
$\one,\two,\three,$ and $(\three+\one)$ representations. The lower
right block of Eq. (\ref{massex}) indicates $\three$ representation
for $\five'$ and $\three+\one$ representation for $\fiveb'$.
Assuming an $S_4$ singlet vacuum for Eq. (\ref{Yukex}), we have
nonvanishing $m_2$ terms but vanishing $m_3$. Anyway, this
illustrates that the number of light SU(5)$'$ quarks are determined
by the choice of the vacuum. Thus, it is possible to find  a six or
seven flavor model of \cite{ISS}. The magnetic phase of the six
flavor model does not have a magnetic gauge group and we must
consider Yukawa couplings only which lead to an infrared free
theory. The magnetic phase of the seven flavor model has the SU(2)
magnetic gauge group but its beta function is positive and the
magnetic phase is again infrared free. Thus, the conclusion on SUSY
breaking studied in the magnetic phase is the desired low energy
phenomenon. In this sense, our model has an ingredient for the GMSB.
Suppose, we have the mass pattern of (\ref{massex}). If $m_{1,2}$ is
near the SU(5)$'$ confining scale, we consider a ten flavor model
down to near the SU(5)$'$ confining scale. So if $m_{1,2}$ are near
the SU(5)$'$ confining scale, some heavy flavors are effectively
removed to be close to a six or seven flavor model and a SUSY
breaking unstable minimum might be a possibility. So we speculate
that in the region $m_{1,2}>\Lambda_h$ an unstable minimum is a
possibility. At the unstable minimum, SU(2)$_W$ is not broken by
hidden sector squark condensates because their values are vanishing
\cite{ISS}.\footnote{But our model is not free from
SU(2)$_W\times$U(1)$_Y$ breaking by $F$-terms of squark condensates
and baryons of the hidden sector. For a more satisfactory model, it
is better to find a SUSY breaking sector being neutral in the SM
gauge group.} For $m_{1,2}\ll\Lambda_h$, an unstable minimum is not
obtained \cite{ISS}. Note that the unification of $\alpha_c$ and
$\alpha_W$ is not automatically achieved as in GUTs because light
$(\one,\two,\one;1;\fiveb',1)_{1/10}$ quarks do not form a complete
representation of a GUT group such as SU(5). Unification condition
must be achieved by mass parameters of the fields surviving below
the GUT scale, and the condition depicted in Fig. \ref{fig:hcoupl}
must be changed accordingly. But we use Fig. \ref{fig:hcoupl} below
just for an illustration.

When SU(5)$'$ confines, there would appear SU(5)$'$ singlet
superfields, satisfying the global (including gauge) symmetries.
Since the remaining six light pairs of  $\five'$ and $\fiveb'$ with
the pattern (\ref{massex}) carry SU(2)$_W,$ SU(2)$_V$ and $Y$
quantum numbers, the composites are formed such that the anomalies
of SU(2)$_W\times$SU(2)$_V\times$U(1)$_Y$ cancel because we know
already that SU(5)$'$ singlet fields of Eq. (\ref{Allspectrum}) do
not carry the SM gauge group anomalies. The remaining six light
pairs of $\five'$ and $\fiveb'$ fields are symmetric under the
interchange SU(2)$_W\leftrightarrow $ SU(2)$_V$, and certainly the
composite leptons will satisfy this symmetry property. Thus, there
is no SM gauge anomaly. In addition, the composite leptons are
standard, i.e. they do not carry exotic charges since the composites
are formed with $(\one,\two,\one;1;\five',1)_{-1/10}$,
$\five'_{2/5}$, $(\one,\one,\two;1;\fiveb',1)_{1/10}$, and
$\fiveb'_{-2/5}$.

\begin{figure}[t]
\begin{center}
\begin{picture}(400,100)(0,0)
{\SetWidth{0.3}  \LongArrow(50,60)(70,60)\LongArrow(53,60)(50,60)
 \Text(60,66)[c]{$r$}}

 \LongArrow(330,60)(350,60)\LongArrow(333,60)(330,60)
 \Text(340,66)[c]{$r$}
\SetWidth{0.3} {\SetWidth{0.9}
 \CArc(60,40)(10,0,180)
 \Curve{(50,40)(50.4,38)(51,35)(54,28)(60,15)(66,28)
 (69,35)(69.6,38)(70,40)} \Text(60,15)[c]{\small$\bullet$}
 \CArc(340,40)(10,0,180)
 \Curve{(330,40)(330.4,38)(331,35)(334,28)(340,15)(346,28)
 (349,35)(349.6,38)(350,40)} \Text(341,15)[c]{\small$\bullet$}

 {\SetWidth{0.3} \Text(210,95)[c]{$R$} \LongArrow(220,95)(270,95)
  \LongArrow(200,95)(150,95)}
 \Curve{(140,30)(150,40)(210,80)} \Text(140,30)[c]{$\bullet$}
  \Text(211,80)[c]{$\bullet$}
 \Curve{(210,80)(258,30)(270,10)} \Curve{(140,30)(200,12)(270,10)}
 \Text(270,10)[c]{$\bullet$}
 }
    \Curve{(200,75)(205,70)(210,70)(213,69)(218,74)}
    \DashCurve{(200,75)(210.5,75)(218,74)}{1.5}
    \Text(210,62)[c]{$\ell_1$}
 \Curve{(180,45)(180.5,46)(183,50)(188,48)(203,50)(207.5,46)(208,44)}
 \Curve{(180,45)(180.5,43)(185,38)(200,40)(203,37)(207.5,42)(208,44)}
     \Text(212,32)[c]{$\ell_0$}

 \Text(60,-5)[c]{(12)}
 \Text(200,-5)[c]{(34)} \Text(340,-5)[c]{(56)}

 {\SetWidth{1}\Text(97,40)[c]{\Large$\otimes$}
 \Text(295,40)[c]{\Large$\otimes$}}
\end{picture}
\caption{The 6d internal space of $T_{1,2,4,7}$ sectors: two pencil
topologies and one triangular ravioli topology. In the (34)-torus,
untwisted string $\ell_0$ and twisted string $\ell_1$ are also
shown.}\label{fig:6Dint}
\end{center}
\end{figure}
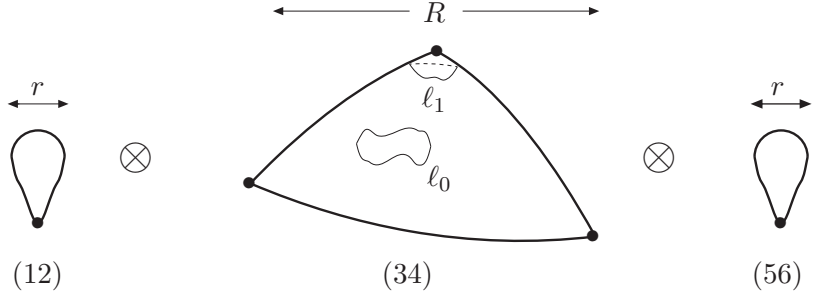

If  $m_{1,2}$ are near the GUT scale, we have a six flavor model,
and the upper dashed line with $-b_j=9$ gives
$\alpha_h\simeq\frac{1}{15}$ for $\Lambda_h=10^{12}$ GeV. If
$m_{1,2}\simeq\Lambda_h$, referring to the lower bold dashed-line of
Fig. \ref{fig:hcoupl}, we have $\alpha_h\simeq \frac{1}{9}$ for
$\Lambda_h=10^{12}$ GeV. These values are large.\footnote{A naive
expectation of the hidden sector coupling, toward lowering the
hidden sector confining scale, is a smaller $\alpha^h_{\rm GUT}$
compared to $\frac{1}{25}$. Because of many flavors,  $\alpha^h_{\rm
GUT}$ turns out to be large.} To introduce this kind of a large
value for the hidden sector coupling constant, we can introduce
different radii for the three tori. In this way, a relatively small
scale, $M_{\rm GUT}\sim 2\times 10^{16}$ GeV compared to the string
scale, can be introduced also via geometry through the ratio $r/R$.
Let the first and third tori are small compared to the second tori
as depicted in Fig. \ref{fig:6Dint}.

If the radius $R$ of the second torus becomes infinite, we treat the
second torus as if it is a fixed torus. Then, one might expect a 6D
spacetime, expanding our 4D spacetime by including the large
(34)-torus. One may guess that the spectrum in $T_1, T_2, T_4$, and
$T_7$ sectors would be three times what we would obtain in $T_{i_0}
(i=1,2,4,7)$. For $T_3$ and $T_6$, the spectrum would be the same
since they are not affected by the Wilson line from the beginning.
But this naive consideration does not work, which can be checked
from the spectrum we presented. If the size of the second torus
becomes infinite, we are effectively dealing with 4d internal space,
and hence we must consider an appropriate 4d internal space
compactification toward a full 6D Minkowski spacetime spectrum. This
needs another set of twisted sector vacuum energies and the spectrum
is not what we commented above. A more careful study is necessary to
fit the hidden sector coupling constant to the needed value. Here we
just comment that in our example SU(5)$'$ is not enhanced further by
neglecting the Wilson line. Even though SU(5)$'$ is not enhanced
between the scales $1/r$ and $1/R$, the SU(5)$'$ gauge coupling can
run to become bigger than the observable sector coupling at the GUT
scale since in our case the bigger group SU(5)$'$, compared to our
observable sector SU(4) group even without the Wilson line, results
between the scales $1/r$ and $1/R$.

The example presented in this paper suggest a possibility that the
GMSB with an appropriate hidden sector scale toward a  solution of
the SUSY flavor problem is realizable in heterotic strings with
three families.

\section{Conclusion}

Toward the SUSY flavor solution, the GMSB from string
compactification is looked for. We pointed out that the GMSB is
possible within a bounded region of the hidden sector gauge
coupling. We find that the hidden sector SU(5)$'$ is the handiest
group toward this direction, by studying the gauge coupling running.
We have presented an example in $\Z_{12-I}$ orbifold construction
where there exist enough number of SU(5)$'$ flavors satisfying the
most needed SM conditions: three observable sector families without
exotics. Toward achieving the needed coupling strength of the hidden
sector at the GUT scale, we have suggested different
compactification radii for the three tori.

\acknowledgments{I thank K.-S. Choi, I.-W. Kim and B. Kyae for
useful discussions. I also thank J.-H. Kim and B. Kyae for checking
the spectrum considered in this paper.
 This work is supported in part by the KRF  Grants, No.
R14-2003-012-01001-0 and No. KRF-2005-084-C00001.
 }



\end{document}